\documentclass[a4paper, 10pt, unnumberedsections, twoside]{LTJournalArticle}

\usepackage[farskip=0pt]{subfig}
\usepackage{lipsum}
\usepackage[acronym]{glossaries}
\usepackage[detect-all=true,per-mode=symbol,per-symbol =/]{siunitx}
\usepackage[capitalise]{cleveref}
\usepackage{tablefootnote}
\usepackage{threeparttable}
\usepackage{booktabs}
\usepackage{xcolor}
\DeclareSIUnit{\x}{\!\ensuremath{\times}}
\DeclareSIUnit\bit{b}
\DeclareSIUnit\gateeq{GE}
\runninghead{}
\footertext{\textit{RISC-V Summit Europe, Munich, 24-28 June 2024}}
\ifx\blind\undefined
    \newcommand{\basilisk}{Basilisk}
\else
    \newcommand{\basilisk}{\textcolor{red}{\scalebox{.6}[1.0]{\textit{BLINDNAME}}}}
\fi

\ifx\blind\undefined
    \newcommand{\basiliskcite}{\cite{basilisk-git}}
\else
    \newcommand{\basiliskcite}{\footnote{Repository omitted for blind review.}}
\fi

\title{{\basilisk}: Achieving Competitive Performance with Open EDA Tools on an Open-Source Linux-Capable RISC-V SoC}

\ifx\blind\undefined
    \author{%
        Phillippe Sauter\textsuperscript{1}\thanks{Corresponding author: \href{mailto:phsauter@iis.ee.ethz.ch}{\tt phsauter@iis.ee.ethz.ch}}, \
        Thomas Benz\textsuperscript{1}, \
        Paul Scheffler\textsuperscript{1}, \
        Zerun Jiang\textsuperscript{1}, \\
        Beat Muheim\textsuperscript{1}, \
        Frank K. Gürkaynak\textsuperscript{1}, \
        Luca Benini\textsuperscript{1,2}
    }
\else
    \author{\textit{Authors omitted for blind review}}
\fi

\newacronym{ml}{ML}{machine learning}
\newacronym{la}{LA}{linear algebra}
\newacronym{qor}{QoR}{quality of results}
\newacronym{soc}{SoC}{system-on-chip}
\newacronym{rtl}{RTL}{register transfer level}
\newacronym{pnr}{P\&R}{place and route}
\newacronym{eda}{EDA}{electronic design automation}
\newacronym{esr}{ESR}{egress SR}
\newacronym{isr}{ISR}{indexed stream register}
\newacronym{cisr}{CISR}{cooperating indexed stream registers}
\newacronym{isa}{ISA}{instruction set architecture}
\newacronym{mv}{M$\times$V}{matrix-vector multiply}
\newacronym{asic}{ASIC}{application-specific integrated circuit}
\newacronym{mac}{MAC}{multiply–accumulate}
\newacronym{fma}{FMA}{fused multiply-add}
\newacronym{drc}{DRC}{design rule check}

\newcommand{\x}{$\times$}

\newcommand{\enumheading}[1]{\newline{\textbf{#1}}\hspace{10pt}}

\ifx\blind\undefined
    \date{
        \vspace{-0.32em}
        \footnotesize\textsuperscript{\textbf{1}}Integrated Systems Laboratory, ETH Zurich \\
        \footnotesize\textsuperscript{\textbf{2}}Department of Electrical, Electronic, and
        Information Engineering, University of Bologna
        \vspace{-0.32em}
    }
\else
    \date{}
    \vspace{1.1cm}
\fi

\renewcommand{\maketitlehookd}{%
	\begin{abstract}
		\noindent We introduce \basilisk, an optimized \gls{asic} implementation and design flow building on the end-to-end open-source Iguana \gls{soc}. We present enhancements to synthesis tools and logic optimization scripts improving \gls{qor}, as well as an optimized physical design with an improved power grid and cell placement integration enabling a higher core utilization. The tapeout-ready version of \basilisk~implemented in IHP's open \SI{130}{\nano\meter} technology achieves an operation frequency of \SI{77}{\MHz} (51 logic levels) under typical conditions, a 2.3\x~improvement compared to the baseline open-source EDA design flow presented in Iguana, and a higher \SI{55}{\percent} core utilization compared to \SI{50}{\percent} in the baseline design. Through collaboration with EDA tool developers and domain experts, {\basilisk} exemplifies a synergistic effort towards competitive open-source \gls{eda} tools for research and industry applications.
	\end{abstract}
}

\begin{document}

\maketitle

\section{Introduction}
Recently, Benz et al.~\cite{tbenz2023iguana} presented and released Iguana, an end-to-end open-source Linux-capable \gls{soc} created with open tools and implemented in IHP's open \SI{130}{\nano\meter} technology.
Iguana is based on the Cheshire platform~\cite{ottaviano2023cheshire} and combines an RV64GC core with a HyperRAM DRAM controller and a rich set of peripherals, completing a Linux-capable system.
The authors released all source files and scripts necessary to run the RTL-to-GDS flow using only freely available open-source tools, enabling others to reproduce and build on their work.

Iguana proved that a completely open-source design flow can be used to implement a Linux-capable \gls{soc}, but the authors did not optimize their design for performance or reduce \gls{drc} violations to a minimal level desirable for tapeout.

In this work, we extend the previous work on Iguana by optimizing the synthesis and the physical implementation of the Cheshire \gls{soc} with open-source \gls{eda} tools; we call the resulting \gls{asic} {\basilisk}.

As was done for Iguana, we avoid simplifications to the original \glsunset{rtl}\gls{rtl} description 
utilizing complex \mbox{SystemVerilog} and instead focus on improving the \gls{eda} tools, their flow scripts, and the overall physical implementation and constraints. %
In the spirit of open-source, we actively foster international and cross-institutional exchange, collecting knowledge and building on existing efforts on cutting-edge algorithms and open \gls{eda} tools. 
We update the \gls{soc} to the newest version of Cheshire, adding new features such as a USB OHCI controller to increase the capabilities of the resulting \gls{asic}. 
Finally, we focus our efforts toward a single, integrated tool flow,
using critical parts of {\basilisk}, such as the FPU or scoreboard,
as realistic benchmarks for all steps in synthesis and \gls{pnr}.

In particular, we present the following contributions:
\begin{itemize}
    \item An extensive study on the state-of-the-art open-source \gls{eda} flow using Yosys for logic synthesis and OpenROAD for \gls{pnr}, 
     with a focus on identifying suboptimal \gls{qor} in individual flow steps. 
    \item Improving existing \gls{eda} tools by communicating the identified tool and flow issues with the maintainers and collaborating on developing solutions.
    \item Providing an optimized open silicon implementation flow, including feedback from various tool authors and domain experts. 
    \item Improving upon Iguana's physical design by optimizing the power grid and adding a manual \gls{rtl} change to facilitate automatic inference of \gls{mac} operations. 
    \item Routability improvements to the standard cells in the IHP \SI{130}{\nano\meter} open PDK.
\end{itemize}

We will further improve {\basilisk} until the second week of May when we plan to tape the resulting \gls{asic} in IHP's \SI{130}{\nano\meter} node.

\section{Synthesis}

Yosys~\cite{wolf2013yosys} is a leading open-source synthesis engine used in open-source \gls{eda} flows. At the time of writing, it offers only limited support for SystemVerilog language constructs; as a result, additional work is needed to convert SystemVerilog, constituting the majority of Cheshire's openly available RTL, to simpler Verilog that Yosys can parse. We use a chain of tools~\cite{tbenz2023iguana} to achieve this: \emph{Morty} merges all dependencies into a single compile context, \emph{SVase} propagates parameters and simplifies some constructs, and \emph{SV2V} converts the remaining SystemVerilog constructs to pure Verilog.    
After elaboration, Yosys first represents the design's structure using high-level constructs and then progressively transforms it to generic standard cells.
Finally, Yosys calls \emph{ABC}, a logic optimization and mapping tool.
To improve \gls{qor}, we improved three distinct aspects of this synthesis chain; their cumulative effects are summarized in \Cref{tab:synth}.
\begin{table}[t!]
    \centering
    \scriptsize{%
        \centering
        \caption{%
            Cumulative synthesis improvements from left to right; large impacts of each step are highlighted.%
        }%
        \label{tab:synth}
        \renewcommand*{\arraystretch}{0.95}
        \begin{threeparttable}
            \begin{tabular}{ccccc} \toprule
                &
                \textbf{Iguana~\cite{tbenz2023iguana}} &
                \textbf{MUX} &
                \textbf{ABC} &
                \textbf{MAC} \\

                \midrule

                \textit{Logic area} &
                \SI{1.8}{\mega\gateeq} &
                \textbf{\SI{1.4}{\mega\gateeq}} &
                \textbf{\SI{1.1}{\mega\gateeq}} &
                \SI{1.1}{\mega\gateeq} \\

                \textit{Timing} &
                \SI{33}{\MHz} &
                \SI{37}{\MHz} &
                \textbf{\SI{71}{\MHz}} &
                \textbf{\SI{77}{\MHz}} \\

                \textit{Logic levels~\tnote{a}} &
                182 LL &
                149 LL &
                \textbf{54 LL} &
                \textbf{51 LL} \\

                \textit{Runtime~\tnote{b}} &
                \SI{5.4}{\hour} &
                \textbf{\SI{2.8}{\hour}} &
                \SI{2.2}{\hour} &
                \SI{2.2}{\hour} \\

                \textit{Peak RAM~\tnote{c}} &
                \SI{217}{\giga\byte} &
                \textbf{\SI{105}{\giga\byte}} &
                \SI{76}{\giga\byte} &
                \SI{75}{\giga\byte} \\

                \bottomrule
                
            \end{tabular}

            \begin{tablenotes}[para, flushleft]
                \item[a] Number of logic gates in longest path
                \item[b] \SI{2.5}{\GHz} Xeon E5-2670
            \end{tablenotes}
        \end{threeparttable}
    }
\end{table}

\textbf{Part-select operations (MUX)}\hspace{10pt}
Yosys versions prior to our improvements (\emph{0.34} and older) use shift operations to represent indexed part-select operations instead of more efficient \emph{block-multiplexer trees}.
Thus, for any part select, a barrel shifter is inferred at elaboration.
This unnecessary generalization of part selects significantly inflates area and logic levels and increases runtime and peak memory usage. 
Logic optimizations in later stages are unable to simplify these shifters to the desired multiplexer trees, impacting \gls{qor}.
\enumheading{Overhaul of the ABC scripts (ABC)}
In cooperation with logic synthesis researchers and ABC developers, we overhaul the ABC script, leveraging Yang et al.'s work~\cite{lazy-synthesis} to improve the \gls{qor} at the cost of minimal additional runtime.
\enumheading{Multiply-accumulate operations (MAC)}
Yosys automatically instantiates optimized implementations of arithmetic operations. 
This also works for \gls{mac} operations, which are implemented as a Booth multiplier followed by an adder.
A more efficient solution is to integrate adders into the \emph{CSA} tree of preceding multipliers, creating \gls{fma} units;
a hand-written \gls{fma} unit reduces our critical path by 9\%.
An update to the existing Booth multiplier transformation automatically inferring \gls{fma} instances is currently in the works.

\section{Place \& Route}
We use \emph{OpenRoad}~\cite{ajayi2019openroad} to implement {\basilisk}'s synthesized netlist. 
Analyzing Iguana's \gls{pnr}, we identify improvement steps mainly in the \emph{\gls{eda} tool flow} (how the individual components of OpenRoad are invoked) and the physical constraints of the \gls{asic}.
We improve the routability of the design by redesigning the power grid;
we reduce the width and increase the count of the power stripes on the top metal layer to ease 
routing congestion underneath the stripes. 
Very dense modules with random routing patterns, such as the boot ROM, were a particular source of issues. As OpenRoad currently only accepts global (as opposed to region- or instance-based) settings, we tune several \emph{hyper-parameters} of the routability-driven global placement engine to improve the placement of dense blocks and get a routable design without \gls{drc} violations.

\begin{figure}[t]
  \centering
  \includegraphics[width=\columnwidth]{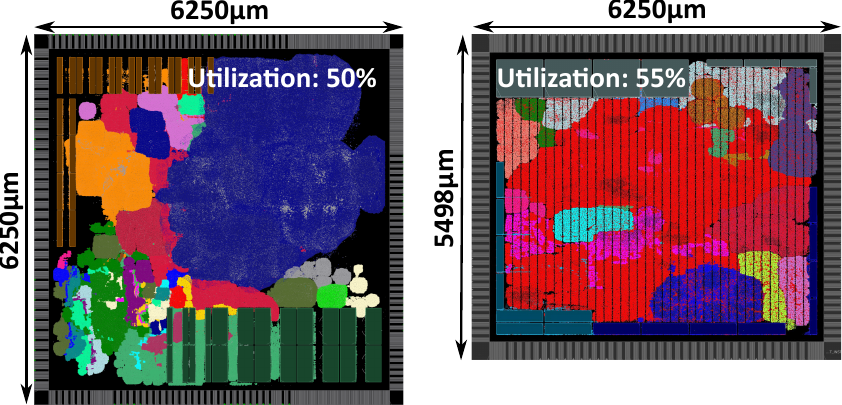}
  \caption{Amoeba view; Iguana (left) and {\basilisk}(right).}    
  \label{fig:die-comparison} 
\end{figure}

\section{Conclusion and Outlook}

Synthesizing and implementing {\basilisk}, we optimize the design's clock frequency by 2.3\x~from \SI{33}{\MHz} to \SI{77}{\MHz} compared to Iguana, reducing the logic area from \SI{1.8}{\mega\gateeq} to \SI{1.1}{\mega\gateeq}.
The physical implementation increases the core utilization from \SI{50}{\percent} to \SI{55}{\percent} while reducing the number of post-routing \gls{drc} violations.
We contribute to improving open-source \gls{eda} tools by reporting and fixing tool issues and by releasing our optimized \emph{flow scripts} and implementation of {\basilisk}~\basiliskcite.
{\basilisk} is not only a theoretical exercise; we will submit our optimized design for fabrication with IHP in their 130nm shuttle run in mid-May.

\printbibliography

\end{document}